\documentstyle[11pt,moriond,epsfig]{article}
\input{psfig.sty}
\begin{document}
%\vspace*{1cm}
\title{QUANTUM PHASE TRANSITIONS IN THE P-SIGE SYSTEM}
\author{ P.T. COLERIDGE, P. ZAWADZKI, A.SACHRAJDA, Y. FENG AND R.L. WILLIAMS }
\address{Institute for Microstructural Sciences,National Research
Council,\\ Ottawa, Ontario, K1A 0R6 , Canada}
\maketitle\abstracts{
The rich variety of phase transitions observed in the strained p-SiGe 
system are considered and compared. It is shown that the integer 
quantum Hall effect transitions, the Hall insulating transition and 
the re-entrant transition into an insulating phase near filling 
factor 3/2 are all very similar and good examples of quantum 
critical phase transitions. The B=0 metal insulator transition also
shows many similarities to these transitions but requires the inclusion 
of an extra impurity scattering term to fully explain the data.
}

%\section{Guidelines}
\section*{Introduction}%\label{subsec:prod}

     The strained p-type SiGe system exhibits a variety of
interesting phenomena. They include: a large and anisotropic
 g-factor \cite{fang92}, the appearance of an insulating phase
\cite{dunford} near filling factor $\nu$ = 3/2 and a
paramagnetic/ferromagnetic spin transition \cite{ssc} between
$\nu$ = 3 and 2 . In addition there is the usual
quantum Hall insulator transition (from filling factor $\nu$ = 1
into the $\nu$=0 insulating phase) and it has also recently been
established \cite{rapid1} that there is a B=0 metal-insulator
transition of the kind observed in high mobility Si-MOSFETs
\cite{kravchenko}. It is argued here that many of these features
are closely related and that the B=0 transition, the $\nu$ = 3/2
transition and the Hall insulating transitions have the same
general character as the standard integer quantum Hall
effect transitions. 

\section*{Samples} 
     The samples used to obtain the results discussed here are
grown by an ultra-high vacuum chemical vapour deposition process
(UHV-CVD). An intrinsic Si layer is followed by a
40nm  $Si_{.88} Ge_{.12}$ quantum well, a spacer layer and a boron
doped silicon layer. The modulation doping ensures the holes from
the ionised acceptors transfer to the quantum well which, because 
the doping is asymmetric, is triangular. The SiGe layer is 
sufficiently narrow that the lattice constant difference between the 
alloy and the pure Si is all taken up by strain. Then the heavy hole 
band, characterised by a  $|M_J|$ = 3/2 symmetry, is well removed 
from other bands. This means the g-factor is large (of order 6) and
depends only on the perpendicular component of magnetic field so,
unlike the more usual situation, the spins cannot be decoupled from the 
orbital motion by tilting the magnetic field.
The large g-factor also means that exchange enhancement of the 
spin-splitting induces a fully spin polarised state \cite{ssc}
 at $\nu$ = 2.
     Typically mobilities are 1-2 m$^{2}$/Vs with quantum mobilities
that are very similar \cite{rapid1} implying the dominant scattering 
potential is short-ranged.

\section*{Integer Quantum Hall transition}

     This system provides one of the few cases where there is a
good theoretical description for the integer quantum Hall
transitions \cite{r4}. For transitions between the  $n_L$ and
$n_L-1$ states the conductivities (expressed in units
of e$^2$/h) are given, following the Chern-Simons boson formulation
\cite{DHL}, in terms of a parameter s with
  
\begin{equation}  
       \sigma _{xx} = 2 \sigma^{pk} s / (1 + s^{2}), \; \; \; \;
                \sigma _{xy} = n_L - s^{2} / (1 + s^{2}).
\label{eq1}  
\end{equation} 
For $n_L$ = 2 the peak value of $\sigma_{xx}$ is  
0.46 and the two components of the conductivity are connected by an 
essentially semi-circular relation (see figure 1). The fact that
$\sigma^{pk}$ deviates from the theoretically expected
universal value \cite{r4} of $1/2$ by only 10 \% 
(the deviation is more usually a factor of order two) is 
attributed to the short-ranged scattering potential. This means the
transport coefficients are not affected by a multiple scattering
momentum weighting term and reflect accurately the underlying
quantum critical phase transition.

\begin{figure} [h]

\vskip 4.2cm

%\special{eps:moriand1.eps x =3cm y=5cm}
\includegraphics{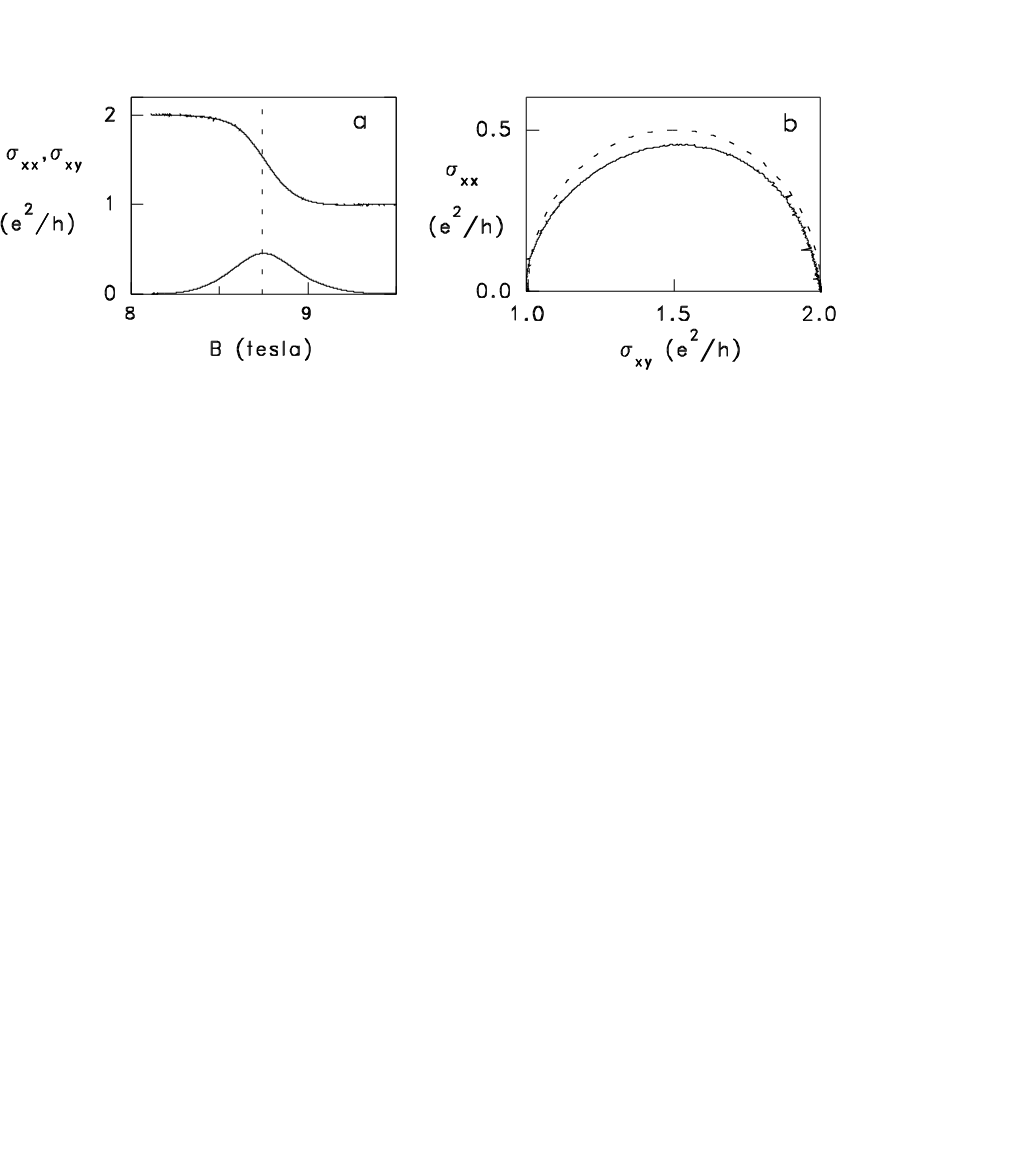}

\caption{Integer quantum Hall transition, $\nu$ = 2 to 1 .
(a) $\sigma_{xx}$ and $\sigma_{xy}$ at T $\approx$ .05K;
(b) $\sigma_{xx}$ versus $\sigma_{xy}$ with semicircle (dashed)
shown for comparison.}

\label{fig1}

\end{figure}

     At low temperatures, and not too far from the critical value $\nu_c$, 
it is found empirically that  
$s     =  \exp [ (\nu_c - \nu) (T_0/T)^{\kappa}]$ ,    
with $\kappa$ close to the theoretically
expected value  \cite{Huckestein} of 3/7. This dependence has also
recently been obtained theoretically \cite{S&W}. The 
same expressions, with $n_L$ = 1 and $\sigma^{pk}$ also
very close to 1/2, explain the  transition into the Hall insulator
state (see figure 2).

\begin{figure} [h]

\vskip 5.0cm
%\special{eps:moriand2.eps}

\includegraphics{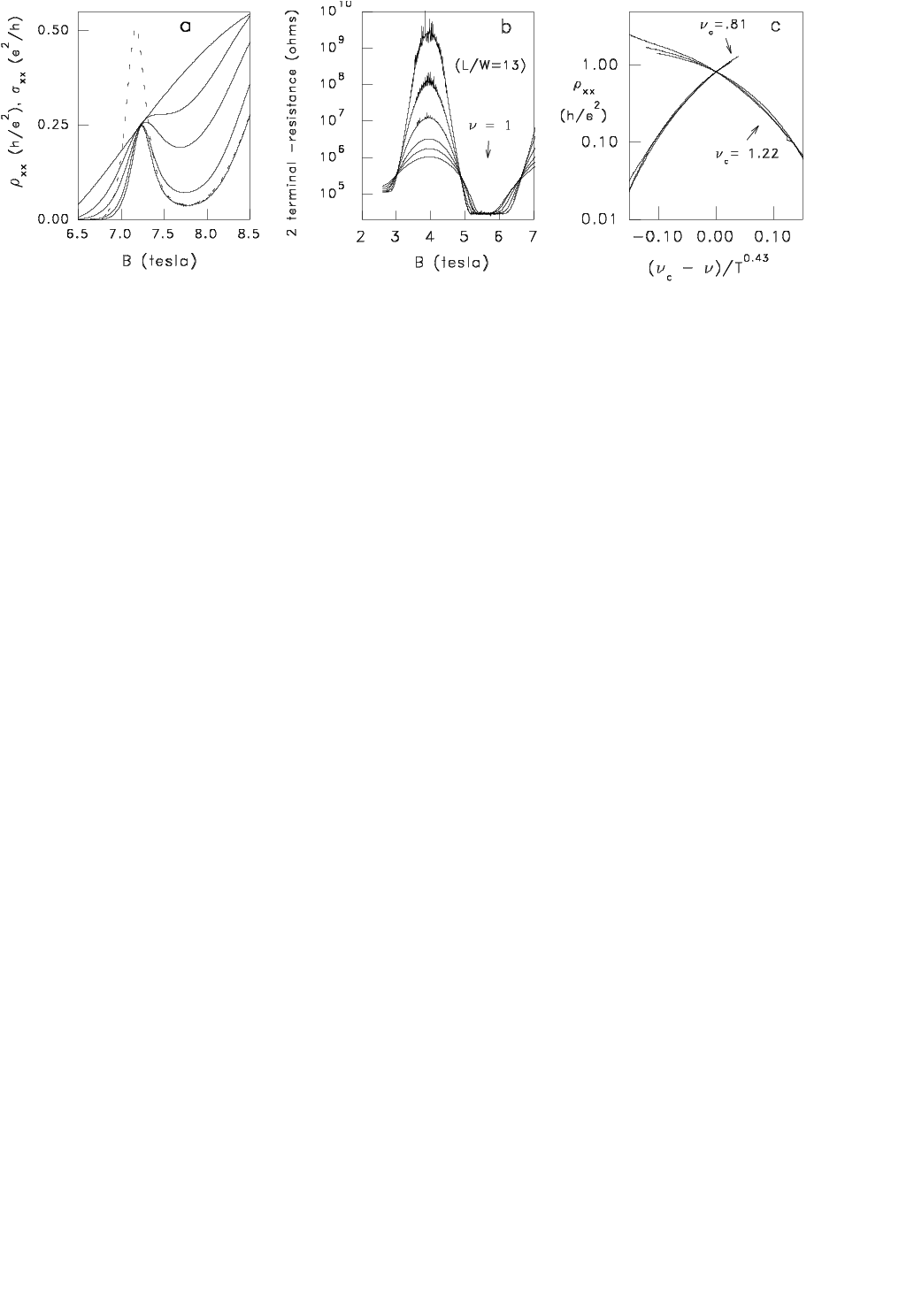}
\caption{(a) Quantum Hall transition, $\nu$ = 2 to 1, for a 
sample (density of 2.6$\times$10$^{15}$m$^{-2}$)
where the $\nu$ = 3/2 insulating phase appears at about 9T.
The dashed line is $\sigma_{xx}$ at 30mK; 
solid lines $\rho_{xx}$ for T = 30 to 1000mK. 
(b) Two terminal resistance of a sample with a density of 
1.4$\times$10$^{15}$m$^{-2}$ , T = 75 - 900mK, showing the
$\nu$ = 3/2 insulating phase with a critical point 
at $\nu_c$=1.22 and the Hall insulating transition 
(with $\nu_c$ = .81). (c) Scaling plot for the 120,270 and 
400mK data from (b).}
\label{fig2}

\end{figure}

\section*{Insulating phase at $\nu \approx$ 3/2}

     An insulating phase near $\nu$ = 3/2 has been reported
\cite{dunford,ssc} in many high mobility p-SiGe samples. While
this often appears to pre-empt and replace the $\nu$ = 2 to 1
quantum Hall transition a careful 
examination of the data on the low field side of the transition 
(see figure 2a) shows it develops from the $\nu$ = 1 quantum 
Hall phase and is re-entrant,  returning to the same $\nu$ = 1 phase 
at higher fields. As shown in figure 2c it is clearly similar
to the Hall insulator transition with the same critical resistivity 
 and temperature scaling. Scaling is poor, however, actually in the 
insulating phase because of the re-entrant character and
close proximity of the two critical points.

\section*{Ferromagnetically polarised spin state}

     It has been demonstrated \cite{ssc}, that 
a paramagnetic/ferromagnetic phase transition, of the
type first predicted by Guiliani and Quinn \cite{G&Q}, occurs
in p-SiGe. This results from exchange enhancement of the spin
splitting, which increases as the system becomes progressively
polarised, leading eventually to a fully polarised spin system at 
$\nu$ =2. The question arises of whether there is a reverse 
transition (back into a paramagnetic state) as the field is
further increased to $\nu$=1.  Analysis of 
activation measurements \cite{ssc}
show that this reverse transition occurs for smaller values 
of the (bare) g-factor and that it is associated 
with the emergence of the insulating phase. This interesting 
correlation, whereby the insulating phase seems to appear only
out of the paramagnetic state, and is suppressed by a
ferromagnetic spin polarisation, is not yet understood.

\section*{B = 0 metal insulator transition}

     Results obtained in a series of p-SiGe samples \cite{rapid1},
where the density was varied by changing the spacer thickness,  show 
the temperature coefficient of the resistivity (at low T) changes 
from negative (ie insulating behaviour) to positive (metallic)
as the density increases. The system exhibits the same B=0 
metal-insulator transition first observed in high mobility Si-
MOSFETs \cite{kravchenko} and also found in p-GaAs 
\cite{pGaAs,Simmons}
and n-AlAs \cite{AlAs}. As in the other systems the 
ratio of the coulomb energy to the Fermi energy is large,
typically 5-10 here. 

     In the insulating phase the resistivity varies as  $\rho_c
\exp[ (T_0/T)^{m}]$ with m of order 0.5, the prefactor $\rho_c$ 
approximately 0.5 h/e$^2$ and T$_0$ varying with 
density. Good scaling behaviour is observed \cite{rapid1} with 
$T_0^m$ proportional to (p$_c$ - p) with p the density and p$_c$ 
the critical density.

\begin{figure} [h]

\vskip 5.1cm

\includegraphics{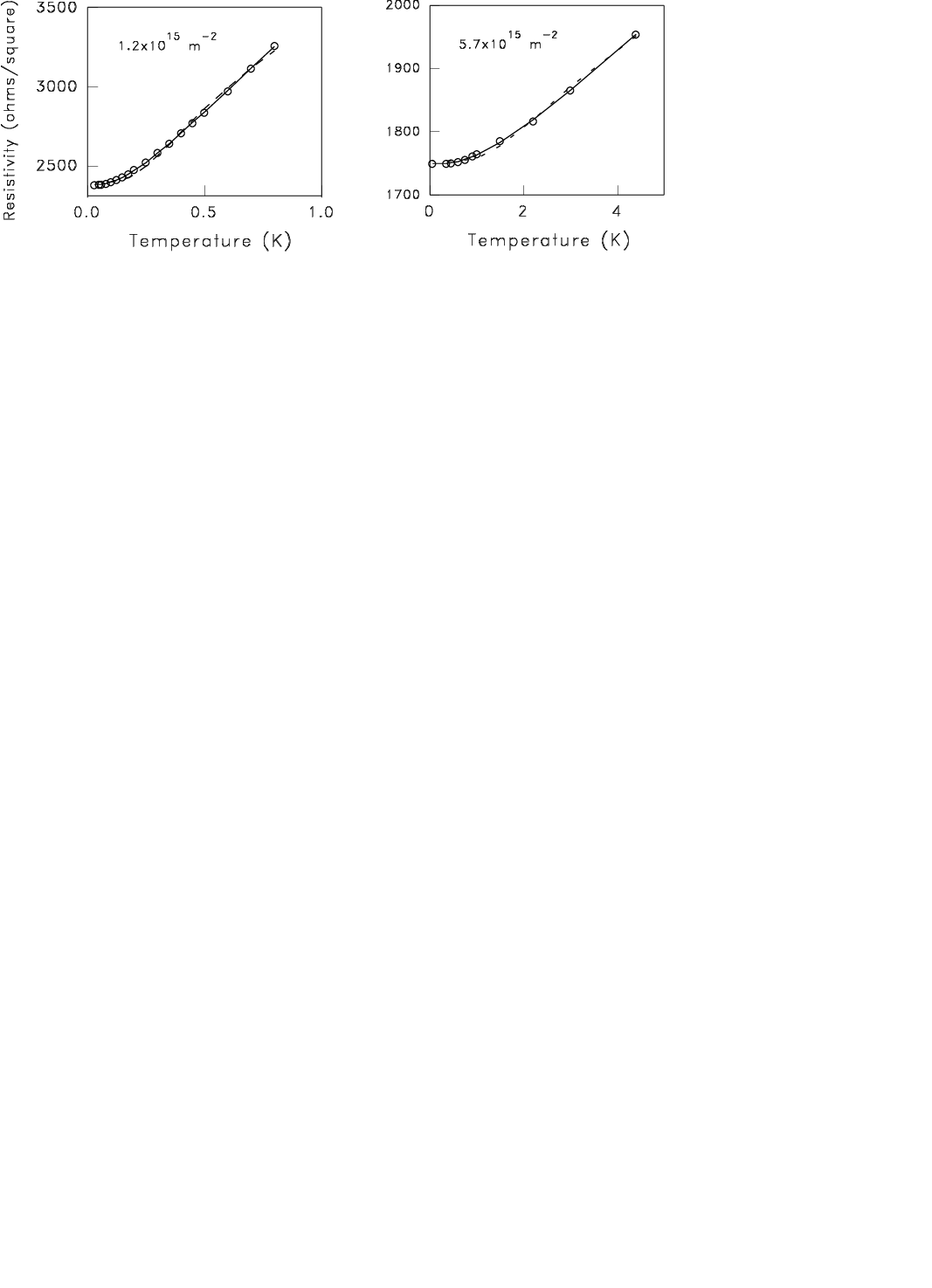}

\caption{ Temperature dependence of $\rho_{xx}$ at B = 0 for two samples
with fits to eqn. 2 for n = 0.4 (solid line) and n = 1 (dashed line).
In the first case $\rho_1$ is of order 0.5 h/e$^2$; in the second .08 
and .02 h/e$^2$ for samples A and B respectively. }
\label{fig4}

\end{figure}

     On the metallic side of the transition the behaviour is a
little more complex. Data is shown in fig.3 for two samples: A,
 relatively close to the critical density (approximately 
1$\times$10$^{15}$m$^{-2}$) and B, deep in the metallic phase. 
In both cases there is a constant background term and
an exponential increase with temperature of the general form 

\begin{equation}
                \rho (T) = \rho _0  + 
   \rho _1 \, \exp [- (T _0 /T)^{n}],         
\label{eq2}
\end{equation}

The same behaviour is also observed in Si-MOSFETs 
\cite{pudalov} and p-GaAs \cite{pGaAs}. In p-SiGe, fitting to
eqn.2 with n=1, gives a prefactor $\rho_1$ that is 
small: this is also the case in p-GaAs. In
the spirit of the argument given by Pudalov \cite{pudalov}, 
this expression can be understood in terms of two 
scattering processes, one leading to a normal impurity 
resistivity, and the other involving another mechanism such 
as activation across a gap. 

     A somewhat better fit to the data in fig. 3 is
obtained for n $\approx$ 0.4. The prefactor $\rho_1$ is 
then of order 0.5h/e$^2$ and there is a strong similarity
with the expression describing the 
insulating behaviour (but with an exponent of the opposite sign). 
Indeed, because $\rho_0$ is small, the variation over a  
range of densities, including the critical value, can be 
described by

\begin{equation}
                \rho (T) = \rho _0  + 
   \rho _c \, \exp [- A \delta_p /T)^{\kappa}]        
 \label{eq3}
\end{equation}
with A a constant, $\delta_p = (p - p_c)/p_c$, 
$\rho_c$ of order h/e$^2$ and 
$\kappa$ about 0.5. Eqn. 3 is essentially that proposed
by Dobrosavljevi\'{c}{\it et al} \cite{DAMC},
to explain the B=0 transition as a quantum critical phase transition, 
but with the addition of the extra impurity scattering term.

\section*{Weak localisation}

     Some support for this dual scattering model comes from 
the magnetic field dependence. There is no obvious evidence of a 
ln(T) term in figure 3 but the low field magnetoresistance 
(figure 4a) has the characteristic shape 
associated with phase breaking of weak localisation by
the magnetic field. At higher fields a positive magnetoresistance
develops typical of the Zeeman interaction term. 

\begin{figure} [h]

\vskip 5.3cm

\includegraphics{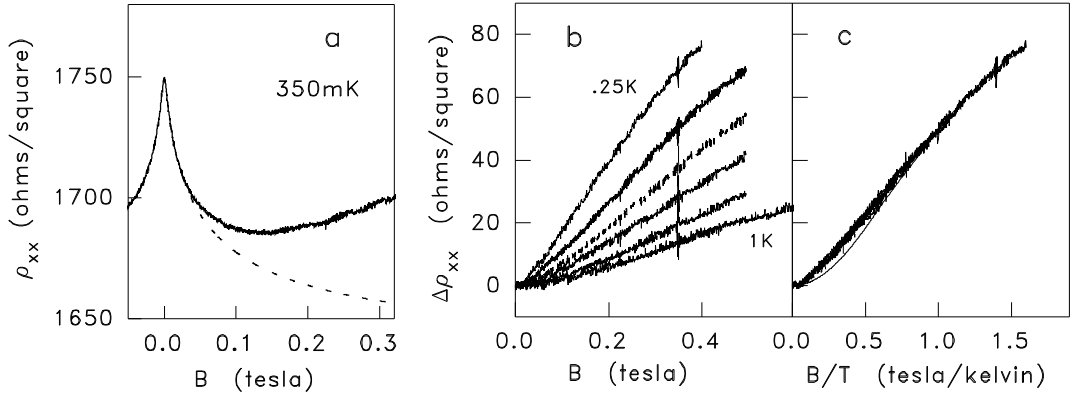}

\caption{ (a) Magnetoresistance, at T = 350mK, for sample B
in fig. 3. The fit to the standard expression  (dashed
line) gives $\tau_{\phi}$ = 46ps and $\alpha$ = .71. (b) Residues
from the weak localisation fits for temperatures of .25, .35, .45,
.6, .75 and 1.0K. (c) Data from (b) plotted against B/T. The solid
line is G(b) (see text) fitted at large B/T.}
\label{fig5}

\end{figure}

 Fitting to the very low B behaviour, using the standard expression 
for the destruction of weak localisation by magnetic field
\cite{L&R},  gives good fits with $\tau_{\phi}$ between 
100 and 3 ps (for temperatures between 0.1 and 1K) and an amplitude 
$\alpha$ of approximately 0.7. The residues from these fits, shown in
figure 4b, are then attributed to the Zeeman interaction term,
 $ \Delta \sigma_2 (B)  =  - ( e^2 / \pi h) ( F^{\ast}/2) G(b)$
(where the function G(b), with b = $g\mu_B B/ k_B T$ is known
\cite{B&D}). Some confidence in this procedure is obtained from the
collapse of this data onto a single curve, when plotted as a function 
of B/T (see fig. 4c). This should be G(b) but 
there are small discrepancies. A fit
at high B/T gives a g-factor of 6.4, consistent with 
the expected value, and an amplitude F$^{\ast}$ of 2.45. 

     The value of F$^{\ast}$ is unphysical (it should be less than 1 )
 but discrepancies of a similar magnitude are also seen in Si-MOSFETs
both for F$^{\ast}$ and for G(b) \cite{B&D}. It is interesting to note
though, that if the derived value of F$^{\ast}$ is used to determine the  
coefficient of the ln(T) dependence, $\alpha p + 1 - 3 F^{\ast}/4$,
it is close to zero (taking the usual value of p=1 for the 
exponent characterising the temperature dependence of $\tau_{\phi}$).
That is, the results appear empirically to be consistent with 
weak localisation although there are some problems with the magnitude of 
the parameters.

     The background, impurity, term therefore appears to behave as expected 
for a standard 2-D system. In particular, it appears to be
weakly localised as T $\rightarrow$ 0 but with a fortuitious
cancellation between the dephasing and Zeeman interaction terms. 

\section*{Summary}

	In the p-SiGe system the Quantum Hall, Hall insulator and $\nu$=3/2
transitions all appear to be good quantum critical phase transitions with
a resistivity that depends exponentially on ($\nu - \nu_c$) and, at least
at low T, scales with a temperature exponent close to 3/7. The B=0 
metal-insulator transition exhibits many of the same quantum critical 
characteristics but appears to involve dual scattering
mechanisms with an additional impurity resistivity, 
at least in the metallic phase, that behaves as a normal 
2-dimensional system including being weakly localised at low temperatures.

\end{document}